# The short-cut network within protein residue networks

Susan Khor
slc.khor@gmail.com
June 5, 2015

**Abstract**

A protein residue network (PRN) is a network of interacting amino acids within a protein. We describe characteristics of a sparser, highly central and more volatile sub-network of a PRN called the short-cut network (SCN), as a protein folds under molecular dynamics (MD) simulation with the goal of understanding how proteins form navigable small-world networks within themselves. The edges of an SCN are found via a local greedy search on a PRN. SCNs grow in size and transitivity strength as a protein folds, and SCNs from "successful" MD trajectories are better formed in these terms. Findings from an investigation on how to model the formation of SCNs using dynamic graph theory, and suggestions to move forward are presented. A SCN is enriched with short-range contacts and its formation correlates positively with secondary structure formation. Thus our approach to modeling PRN formation, in essence protein folding from a graph theoretic view point, is more in tune with the notion of increasing order to a random graph than the other way around, and this increase in order coincides with improved navigability of PRNs.

**Introduction**

A small-world network (*SWN*) is a set of inter-connected nodes whose average path length increases logarithmically with the number of nodes in the network, and whose clustering coefficient is larger than expected from a comparable Erdos-Renyi random graph [1]. The coordinates of crystallized proteins reported in the PDB [2] have been used to construct networks of interacting amino acids within proteins. We call such a network, built per the Methods section, a Protein Residue Networks (*PRN*). The SWN within protein molecules has held a special interest for protein scientists who use its properties to identify functional (e.g. binding and nucleation) sites, and to map the communication network within proteins [3-7]. However, modeling the formation of the SWN within proteins is still a challenge for those seeking an algorithmic understanding of protein folding logic. Most models of SWN evolution to date have a fixed number of nodes and edges and the edges are rewired in some probabilistic fashion to attain some pre-determined goal(s), e.g. clustering, path length, assortativity, edge length distribution or navigability [8-12]. In contrast, the formation of SWNs within proteins is due to the distances between a protein's amino acid molecules (the nodes) as they move in response to each other within a substrate in three dimensional Euclidean space subject to constraints of the protein backbone. And unlike the so-called power-law networks, the number of links per node in a PRN is limited due to the excluded volume argument.

Furthermore, PRNs not only possess SWN structure, they are also navigable small-worlds [13]. Small-world networks are navigable if short paths can be found via a local search, for instance passing a message to an acquaintance who is, one believes from one's limited knowledge of the network's topology, closer to the intended recipient. Kleinberg showed that not all graphs having SWN structure are navigable, and that linking which respects the geometry of the metric space plays a crucial role in efficient local search [14].

We propose that a sparser but more volatile sub-network of a PRN called the short-cut network (*SCN*) makes a suitable object of study for modeling the evolution of navigable SWNs in proteins. A SCN comprises a subset of PRN edges that function as short-cuts in the course of a search by the EDS algorithm (see Methods section for description of EDS). Previously [13], we reported that the number of links in an SCN is about twice its number of nodes, and noted that the links of a SCN are dominated by short-range contacts (links that connect nodes not more than 10 residues apart on the protein sequence [15]), have high centrality (are traversed by many paths), impacts average path length significantly, and are distinct from other PRN links in other ways.

In this paper, we describe the dynamical aspects of SCNs observed in a Molecular Dynamics (MD) simulation of the 2EZN protein unfolding from its native state [16-18], and the HP-35 NLE NLE protein (Villin Headpiece modified for fast-folding, PDB: 2F4K) folding from various denatured states [19] (details in the Methods section, Fig. SM1). We find that a (well-formed) SCN grows as a protein folds to span almost all the nodes of its PRN. In keeping with the role SCNs play in the small-world communication of PRNs, the set of links that make up an SCN undergo significantly more changes (additions and deletions) during protein folding than other PRN links. Despite this volatility in short-cut edge sets, several patterns could be discerned and it is these patterns that could provide insight into the formation of navigable SWNs in PRNs. Moreover, we observed significant differences between SCNs from "successful" and "unsuccessful" MD simulations. The SCNs from "successful" MD trajectories were better formed on average, and SCN well-formness correlates strongly with the presence of native secondary structures and other desirable properties.

**Results**

*SCN size and growth*

The number of short-cut edges |*SC*| increases as the 2EZN protein folds in the MD simulations. Native state (6250) PRNs have significantly larger short-cut sets than non-native state (6251…6258) PRNs (Fig. 1a). The average number of short-cuts over all PRNs from run 6250 is about 2*N* (*N* is the number of nodes in a PRN; for 2EZN $N=101$), but it is only about 1.5*N* for runs 6251…6258. While the SCNs span almost all the nodes of their PRNs, they may not be connected. We are interested in the largest connected

component of a SCN, denoted *gSCN*. The number of nodes in a gSCN is largest for native-state PRNs (6250 in Fig. 1b), and hovers around 95% for non-native state PRNs. Therefore, as the 2EZN protein folds, the gSCN grows in both edges and nodes. The existence of a gSCN that covers many more nodes than all other components in the SCN is not assured given that the 2EZN SCNs have significant levels of clustering or transitivity (Fig. 2a).

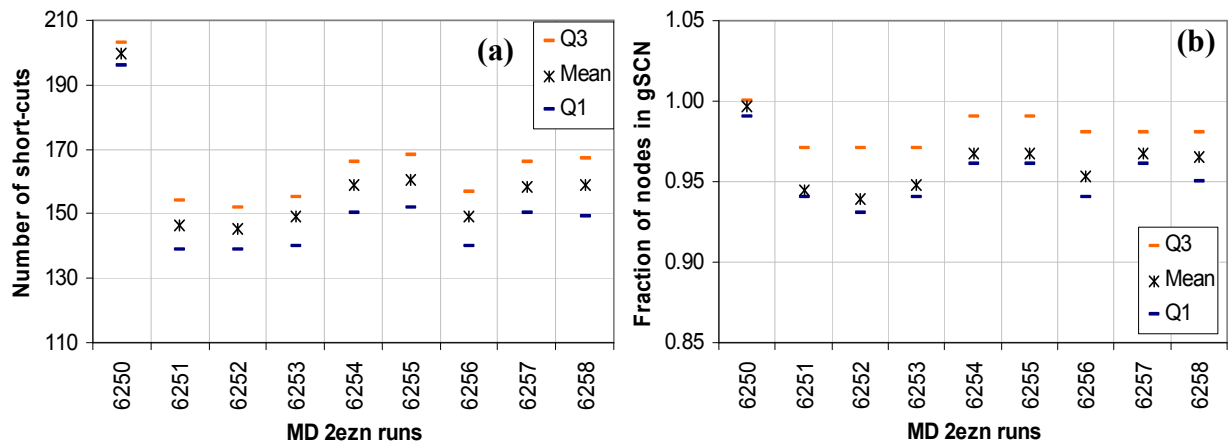

**Fig. 1 (a)** Native state PRNs (6250) have significantly larger short-cut sets than non-native state PRNs (6251…6258). **(b)** The largest component of a SCN (gSCN) spans almost all the nodes of a PRN. gSCN node coverage averages around 95% in non-native state PRNs and is almost 100% in native state PRNs (6250). Q1 and Q3 denote the first and third quartiles respectively.

*SCN transitivity, path diversity and path-pair stretch*

SCN transitivity is the average multiplicity of edges in a SCN (evaluated solely within the SCN). The multiplicity of an edge *e* is the number of distinct triangles that passes through *e* [20]. SCN transitivity is stronger when average edge multiplicity is larger. SCN transitivity increases in strength as the 2EZN protein folds in the MD simulation (Fig. 2a). The native state PRNs have significantly stronger SCN transitivity than the non-native PRNs. SCN transitivity strength influences path diversity (Fig. 2b) and average path length via path-pair stretch (Fig. 2c).

Proteins are fairly robust to random attacks according to mutagenesis studies [5], possess alternative pathways or redundant links between critical sites [4, 21], and can modulate allosteric pathways to fit conditions at hand [22]. Hence the existence of alternate pathways (*edge independent* paths) with minimal difference in length (small *stretch* factor) is an essential feature for PRNs. Let *p* be a path from node *u* to node *v*, and *q* be a path from *v* to *u*. Paths *p* and *q* make a path-pair. *p* and *q* is an edge independent or edge disjoint path-pair if and only if *p* and *q* have no edges in common. A network that supports the existence of many edge independent path-pairs is *path diverse*. The stretch of a path-pair is the absolute difference between the lengths of its constituent paths. Path diversity and stretch is a function of network topology and search algorithm. Given a PRN and no special effort to create diverse paths, our EDS

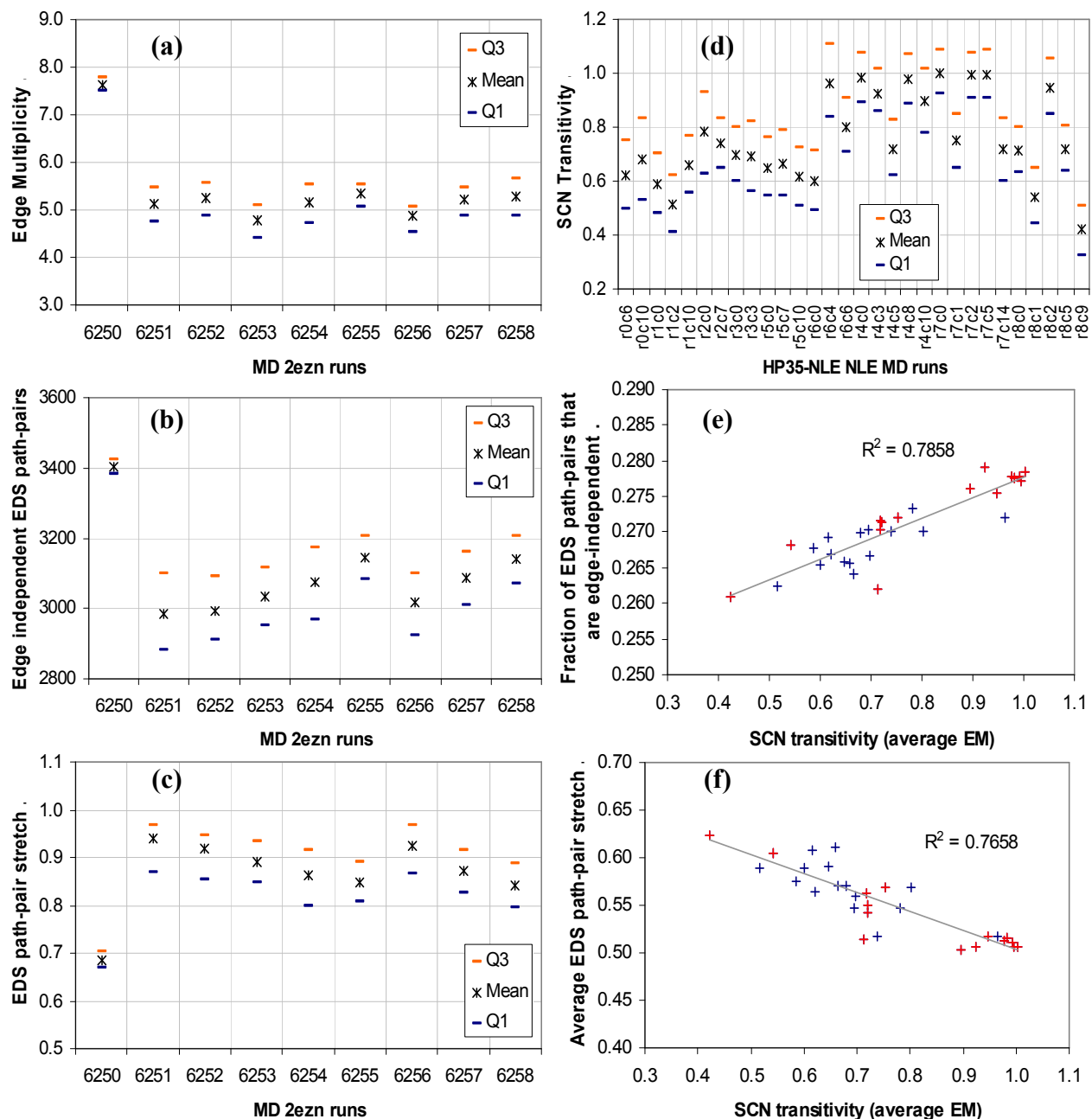

**Fig. 2 (a)** SCN transitivity grows stronger as the 2EZN protein folds. Native state PRNs (6250) have significantly stronger transitivity than non-native state PRNs (6251…6258). **(b)** A PRN with a larger number or a higher fraction of EDS path-pairs that are edge independent is more path diverse. Path diversity increases as the 2EZN protein folds. **(c)** The stretch of an EDS path-pair is the absolute difference between the lengths of the EDS path-pair from $u$ to $v$ and from $v$ to $u$. EDS path-pair stretch decreases as the 2EZN protein folds. Plots (d-f) report observations from the Villin MD dataset. **(d)** SCNs from the "successful" trajectories (r4c0…r8c9) have significantly stronger transitivity than SCNs from the "unsuccessful" trajectories (r0c6…r6c6). **(e)** SCN transitivity correlates strongly and positively with path diversity. **(f)** SCN transitivity correlates strongly and negatively with EDS path-pair stretch. In (e) and (f), the red and blue crosses denote successful and unsuccessful MD runs respectively. The two red crosses in the left half of plots (e) and (f) correspond to runs r8c1 and r8c9 which have noticeably weaker SCN transitivity. These two exceptional cases are discussed in the text.

algorithm finds significantly more edge-independent path-pairs than a Breath-First-Search (BFS) algorithm. All BFS path-pairs, even the edge-independent ones, will have zero stretch. Path diversity increases while path-pair stretch decreases as the 2EZN protein folds. PRNs of native state configurations (6250) have significantly more edge independent EDS path-pairs (Fig. 2b) and significantly smaller EDS path-pair stretch (Fig. 2c) than the PRNs of non-native state configurations (6251…6258).

The relationship between SCN transitivity (structure), and path diversity and stretch (function) observed in the 2EZN MD simulation is also observable with the Villin MD dataset. The PRNs from the "successful" Villin MD runs have SCNs that are significantly more strongly transitive on average than the PRNs from the "unsuccessful" MD runs (Fig. 2d). Accordingly, EDS path-pairs from the "successful" MD runs are significantly more edge independent (Fig. 2e), and suffer significantly smaller average stretch than EDS path-pairs from the "unsuccessful" MD runs (Fig. 2f). These differences, together with the significant difference in SCN size (Table 1), suggest that SCNs are better formed in the "successful" than in the "unsuccessful" Villin MD trajectories.

**Table 1 Significant differences between "successful" and "unsuccessful" Villin MD runs.** SCNs of "successful" runs are better formed in the sense that they have more edges, larger gSCNs, stronger transitivity and more native short-cuts. Further, their PRNs have significantly more alpha structure residues. Well-formed SCNs confer functional (path diversity and stretch) and developmental (robustness) advantages. *one-sided, unpaired. Sample size: 15 "successful" runs and 15 "unsuccessful" runs. $N$=35 for Villin PRNs.

| Metric | "successful" runs avg ± std. dev. | "unsuccessful" runs avg ± std. dev. | t-test* p-value |
|---|---|---|---|
| Number of short-cuts | 53.3627 ± 2.1543 | 51.4073 ± 1.8910 | 0.0067 |
| Number of nodes in the largest SCN component | 33.8007 ± 0.5008 | 33.1513 ± 0.6717 | 0.0029 |
| SCN transitivity (average edge multiplicity) | 0.8202 ± 0.1809 | 0.6845 ± 0.1071 | 0.0101 |
| Fraction of EDS path-pairs that are edge-independent | 0.2727 ± 0.0059 | 0.2680 ± 0.0031 | 0.0070 |
| EDS path-pair stretch | 0.5377 ± 0.0394 | 0.5682 ± 0.0279 | 0.0073 |
| Fraction of native short-cuts | 0.6043 ± 0.1521 | 0.4728 ± 0.0815 | 0.0037 |
| Number of alpha structure residues (DSSP) | 15.3033 ± 5.6177 | 10.6982 ± 3.1652 | 0.0056 |
| Fraction of snapshots with > 35 native short-cuts | 0.4892 ± 0.4237 | 0.1085 ± 0.1409 | 0.0021 |

*SCN transitivity, native contacts and secondary structure*

There are two runs (r8c1 and r8c9) which have notably weaker SCN transitivity (Fig. 2d) but are classified as "successful" in [19]. In terms of the fraction of native SCN contacts and the number of residues with alpha secondary structure according to DSSP [23], these two clones are laggards compared with the other "successful" clones (Table 1, Fig. 3a).

Native SCN contacts are the SCN edges of the 2F4K PRN generated from the PDB structure. Let the native SCN be $SCN_0$. The fraction of native SCN contacts in a SCN at step $t$ ($SCN_t$) is the number of edges $SCN_t$ has in common with $SCN_0$ divided by the number of edges in $SCN_0$, which is 55. The alpha structure residues for a PRN were obtained by running DSSP on the PRN's coordinates file, and counting the number of residues marked 'H', 'I' or 'G'. Despite having these two "outlier" runs (r8c1 and r8c9), the "successful" Villin MD trajectories register on average a significantly larger fraction of native SCN contacts, and significantly more alpha structure residues than the set of "unsuccessful" Villin MD runs (Table 1).

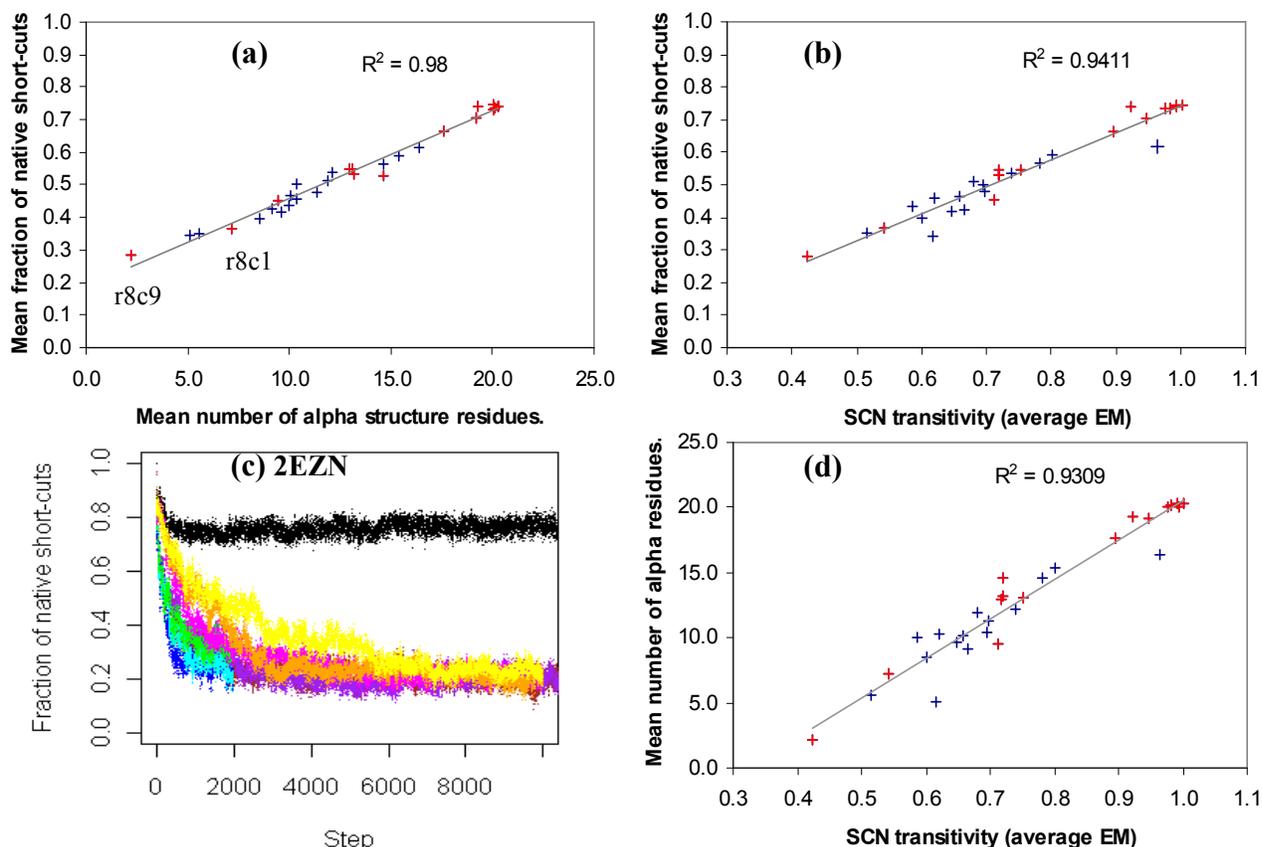

**Fig. 3 (a, b, d) Relationship between fraction of native short-cuts, alpha structure residues and SCN transitivity observed with the Villin MD dataset.** Native short-cuts are the short-cut edges in SCN0, the SCN of the 2F4K PDB structure. Mean fraction of native short-cuts for a trajectory is the fraction of edges in a SCN that are native short-cuts averaged over all SCNs (steps) in the trajectory. 2F4K comprises three helical structures (Fig. SM1). **(a)** The mean fraction of native short-cuts is strongly and positively correlated with the mean number of alpha structure residues. Except for r8c9 and r8c1, the "unsuccessful" trajectories tend to cluster in the lower left quadrant of the plot. **(b)** The mean fraction of native short-cuts is strongly and positively correlated with SCN transitivity. **(c)** The fraction of native short-cuts in each SCN increases steadily as the 2EZN protein folds, and holds steady for native dynamics. **(d)** SCN transitivity is strongly and positively correlated with mean number of alpha residues.

Native contacts play a dominant role in protein folding and the fraction of native contacts was found to be a suitable folding coordinate even for all-atom MD simulations [24]. The fraction of native SCN contacts increases steadily as the 2EZN protein folds under MD simulation (Fig. 3c), implying that SCN formation is (free) energetically downhill. The "successful" Villin MD trajectories have significantly more SCNs whose short-cuts are at least 65.45% native (Table1, Fig. SM2). SCN transitivity correlates positively and strongly with both fraction of SCN native contacts (Fig. 3b) and number of alpha residues (Fig. 3d). This is expected since short-cuts are dominated by short-range links [13], and short-range links are found mainly within secondary structures. These strong correlations, coupled with the significant differences in Table 1, suggest that the characteristics of SCNs could offer a more computationally efficient way to distinguish between protein folding pathway possibilities [25].

*SCNs and different folding pathways*

Both the B1 immunoglobulin binding domain of streptococcal protein G (PDB: 1GB1) and the B1 immunoglobulin binding domain of peptostreptococcal protein L (PDB: 2PTL residues 17 to 78) comprise a four-strand β-sheet and a single α-helix (Fig. 4). In spite of their identical secondary structure make up and very similar tertiary structures, this much studied protein pair folds differently, and the difference in their folding pathways is discernable even at the mesoscopic level [25]. Yang and Sze [25] predicted protein folding pathways for several proteins by first transforming a protein sequence into a ring of fully folded secondary structure elements (SSEs) mostly interspersed with turns, and then iteratively pairing the SSEs or combination thereof, which produces the lowest free energy in each step. They support their predictions with experimental results reported in the literature.

Let 1S, 2S, 3S and 4S denote the four strands of the β-sheet, 1H denote the α-helix and $n$T denote the $n^{th}$ turn. In this schema (data available online at faculty.cs.tamu.edu/shsze/ssfold), the 1GB1 protein sequence transforms to ⟨1S, 1T, 2S, 2T, 1H, 3T, 3S, 4T, 4S⟩, and the 2PTL protein sequence is expressed as ⟨1S, 1T, 2S, 2T, 1H, 3T, 3S, 4T, 4S, 5T⟩. The predicted fold order for 1GB1 is (((3S4S)1H)1S)2S), which means 3S and 4S combines first, then the combination of 3S and 4S combines with 1H, and so on. The predicted fold order for 2PTL is ((((1S2S)(3S4S))1H), which means the N-terminal β-hairpin ((1S2S)) forms first, followed by the C-terminal β-hairpin ((3S4S)), then the four-strand β-sheet (((1S2S)(3S4S))) and finally the α-helix gets combined. These predictions concur, in the main, with experimental observations that folding in 1GB1 starts with the formation of the C-terminal β-hairpin, while in 2PTL the N-terminal β-hairpin forms first.

We find that differences in the folding pathways of 1GB1 and 2PTL can show up as differences in SCNs. The SCNs of 1GB1 and 2PTL are more dissimilar than their PRNs (Figs. 4A & 4B upper left triangle). Unlike 1GB1, 2PTL has no short-cuts between (1S2S) and (3S4S). However, these short-cuts can appear at an intermediate stage if 2PTL follows 1GB1's fold order (Fig. 4H). Similarly, spurious (relative to native SCN) short-cuts appear if 1GB1 follows a different fold order, e.g. combining 2S before 1S (Fig. 4G). Our results were obtained by adding inter-SSE links and links involving turn residues, in stages according to a fold order, to a PRN initially built with intra-SSE links only. Links involving turn residues are added if both SSEs adjacent to the turn are combined. SCNs for partially formed PRNs were obtained in the same way as fully formed PRNs, i.e. by finding EDS paths between all possible node-pairs.

**1GB1  ((((3S4S)1H)1S)2S)**    **2PTL  (((1S2S)(3S4S))5H)**

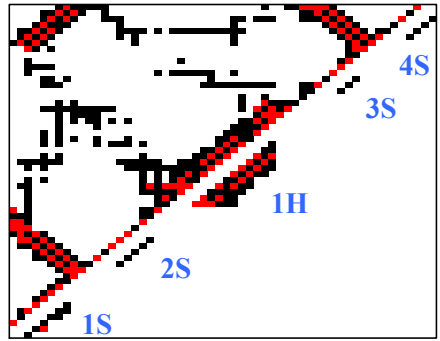  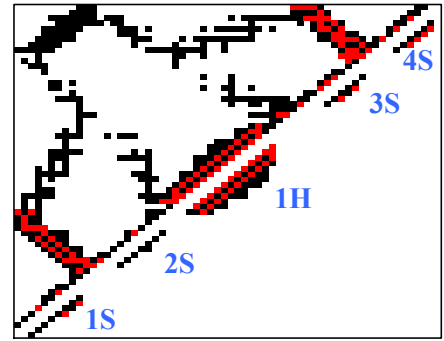

Upper-left triangle: Contact map of fully formed PRN (black), and its SCN (red).

Lower-right triangle: Initial PRNs comprise only intra-SSE links. The SSEs are 1S, 2S, 3S, 4S and 1H. PRN links are in black, SCN edges are in red.

A    B

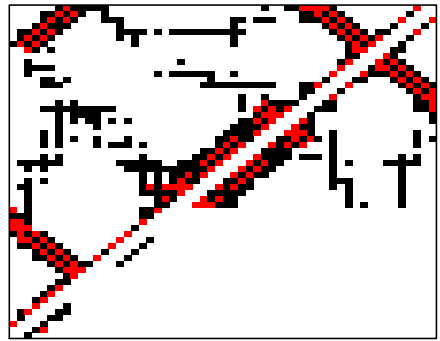  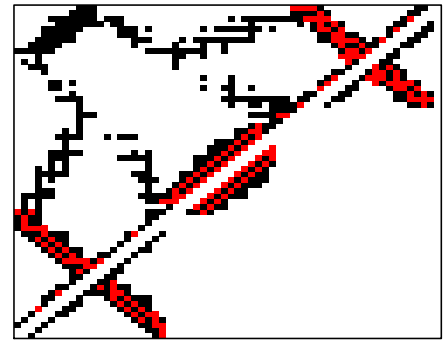

Upper-left triangle: Contact map of fully formed PRN (black), and its SCN (red).

Lower-right triangle: PRN and SCN after two fold events.
**C** 1GB1: ((3S4S)1H)
**D** 2PTL: (1S2S)(3S4S)

C    D

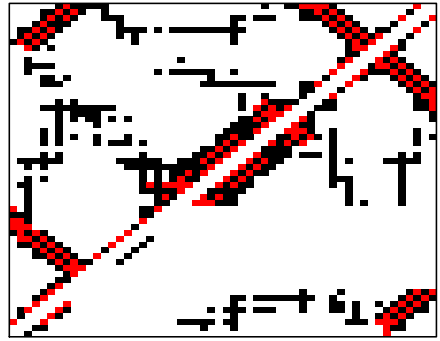  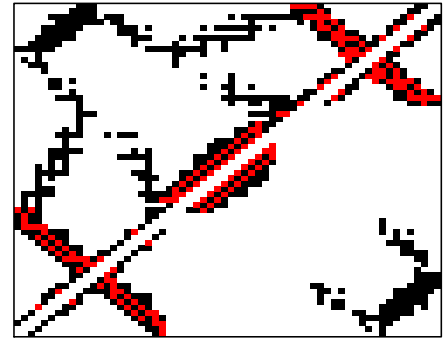

Upper-left triangle: Contact map of fully formed PRN (black), and its SCN (red).

Lower-right triangle: PRN and SCN after three fold events.
**E** 1GB1: (((3S4S)1H)1S)
**F** 2PTL: ((1S2S)(3S4S))

E    F

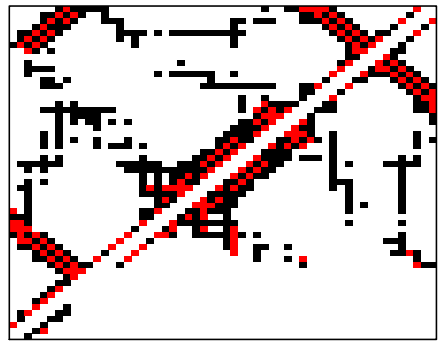  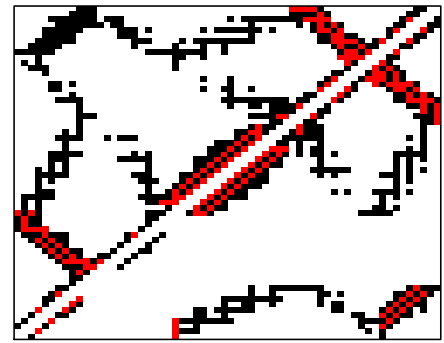

Upper-left triangle: Contact map of fully formed PRN (black), and its SCN (red).

Lower-right triangle: PRN and SCN after a misfolding event(s):
**G** 1GB1: (((3S4S)1H)2S)
**H** 2PTL: (((3S4S)1H)1S)

G    H

**Fig. 4** Fully and partially formed PRNs and SCNs of 1GB1 and 2PTL.

*Volatility of SCN edges*

To preserve or increase path diversity and to maintain or reduce path stretch, the formation and destruction of short-cut edges need to be responsive to the changes in paths and their lengths. Compared to the set of edges that are not short-cuts, the set of short-cut edges is significantly more volatile (Fig. 5). The diagram on the right of Fig. 5 summarizes the possible transitions between the three different states a contact map entry can be in: 0 is the non-edge state, 1 is the non-short-cut edge state and 2 is the short-cut state. In an MD run, zero or more of these transitions can happen from one step to the next.

Non-edges are most likely to remain in their current state as non-edges in the next step (*d00*). Since PRNs are sparse, it is unsurprising that almost all zero entries in the contact map (adjacency matrix of a PRN) at step *t* remain zero at step *t*+1. Except for 6250, about 80% of non-short-cut links remain as such in the next step time (*d11*). Except for 6250, about 70% of short-cut links remain as such in the next step time (*d22*). This means that, except for the native state PRNs (6250), a short-cut edge is significantly more likely to transition out of its current state, i.e. become a non-short-cut edge (*d21*) or a non-edge (*d20*) in the next step, than a non-short-cut edge. This significantly higher percentage of state change is why short-cut edges are considered more *volatile* than non-short-cut edges. Run 6250 is a native dynamics simulation. Expectedly, its *d11* and *d22* values are both significantly larger than the other non-native runs.

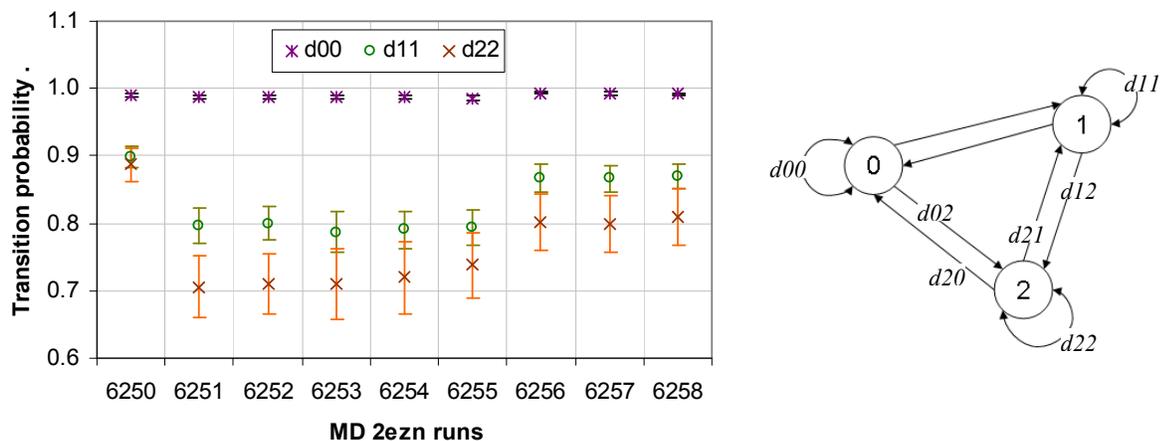

**Fig. 5 Short-cut edges are significantly more volatile than non-short-cut edges**. *d00* denotes non-links that persist from one step to another. *d11* denotes non-short-cut links at step *t* that remain a non-short-cut link at step *t*+1. *d22* denotes short-cut links at step *t* that remain a short-cut link at step *t*+1. Error-bars denote standard deviation about the mean. Right: all the possible edge transitions.

*SCN robustness*

From Fig. 1 we observed that a SCN faces the task of node percolation on its PRN, i.e. growing its largest connected component (gSCN). Serrano and Boguñá [26] found that with degree distribution and clustering coefficient values being equal, networks with strong transitivity are more resilient to random edge removals than weakly transitive networks. The reduction in size of the giant component as edges are removed uniformly at random actually slows down in strongly transitive networks as the probability of removing a random edge $q$ increases. Conservatively, this effect is pronounced for $q > 0.2$ (Fig. 4 in [26]).

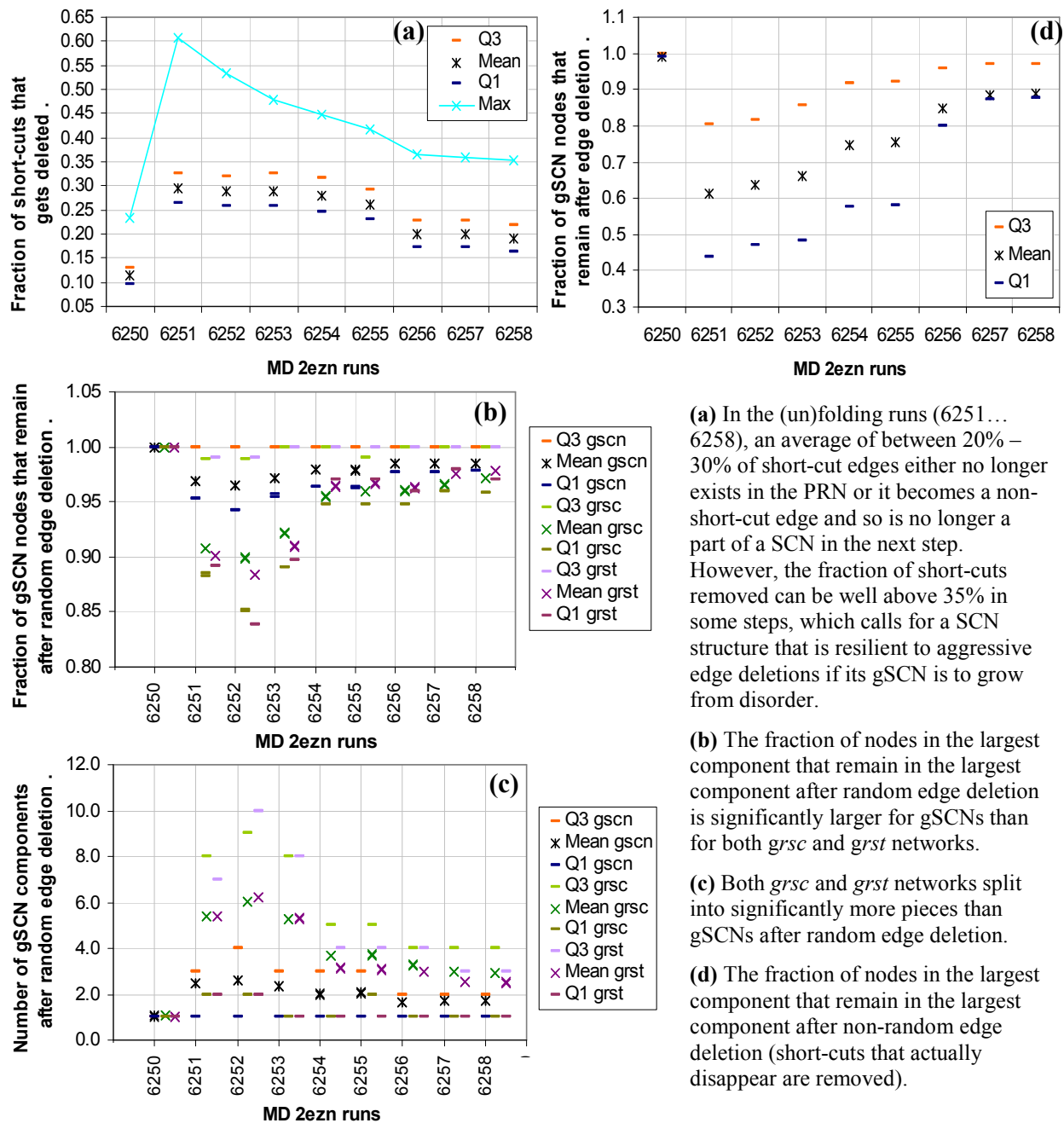

(a) In the (un)folding runs (6251…6258), an average of between 20% – 30% of short-cut edges either no longer exists in the PRN or it becomes a non-short-cut edge and so is no longer a part of a SCN in the next step. However, the fraction of short-cuts removed can be well above 35% in some steps, which calls for a SCN structure that is resilient to aggressive edge deletions if its gSCN is to grow from disorder.

(b) The fraction of nodes in the largest component that remain in the largest component after random edge deletion is significantly larger for gSCNs than for both g*rsc* and g*rst* networks.

(c) Both *grsc* and *grst* networks split into significantly more pieces than gSCNs after random edge deletion.

(d) The fraction of nodes in the largest component that remain in the largest component after non-random edge deletion (short-cuts that actually disappear are removed).

**Fig. 6 Robustness of the largest SCN component (gSCN) to (random) edge deletions relative to *grsc* and *grst*.**

In the 2EZN MD dataset, on average between 20% – 30% of short-cut edges disappear in a step in non-native runs (6251…6258 in Fig. 6a). A short-cut edge disappears when it either no longer exists in the PRN (*d20* transition) or it becomes a non-short-cut edge and so is no longer a part of a SCN (*d21* transition). The fraction of short-cuts removed can be well above 35% in some steps, which calls for a SCN structure that is resilient to aggressive edge deletions. And indeed, the more strongly transitive gSCNs show significantly more resilience to random edge removals than the weakly transitive *grsc* and g*rst* networks (see Methods section for details). *grsc* (*grst*) denotes the largest connected component of a *rsc* (*rst*) network. The weak transitivity of a *rsc* network is expected since it is essentially a random graph, i.e. its edges are selected uniformly at random from PRN edges and so are not expected to be correlated. The weak transitivity of a *rst* network is also expected since it is constructed from the union of random spanning trees. The fraction of nodes in the largest component that remain in the largest component after random edge deletion is significantly larger for gSCNs than for both g*rsc* and g*rst* networks (Fig. 6b). After random edge deletion, both *grsc* and *grst* networks fracture into significantly more pieces than gSCNs (Fig. 6c). The native configurations in the MD simulation have significantly stronger SCN transitivity than non-native PRNs (Fig. 2a), and this strength is reflected in the results in Fig. 6 where the size of native gSCNs (6250) is almost immune to both random and non-random edge deletions.

Let $q_t$ be the fraction of edges deleted from gSCN$_t$ (gSCN at step *t*); and let *grsc*$_t$ and *grst*$_t$ be respectively, the largest component of the *rsc* and *rst* networks associated with PRN$_t$. Fig. 6a gives the *q* value averaged over all steps in a run. For results reported in Figs. 6b & 6c, $q_t$ of the edges in gSCN$_t$, *grsc*$_t$ and *grst*$_t$ were chosen uniformly at random for removal in each step *t*, and the results are averaged over all steps in a run. In Fig. 6d, $q_t$ of the edges in gSCN$_t$ is also removed but these edges are the actual short-cut edges that disappear at step *t*+1. The fraction of nodes in the largest component that remain in the largest component after this non-random edge deletion is significantly smaller than when the same number of random edges is removed from gSCN. Thus, the deleted short-cut edges are not a random edge set, but possibly a deliberate stressor that encourages a sufficiently resilient gSCN to grow.

*SCN function influences its formation*

EDS path-pairs for each PRN were partitioned into four types: *00* denotes edge independent path-pairs with zero stretch, *01* denotes edge independent path-pairs with positive stretch, *10* denotes path-pairs that are not edge independent with zero stretch, and *11* denotes path-pairs that are not edge independent with positive stretch. Single-edge paths were excluded from this analysis so that the results for type *10* are not artificially inflated. The distribution of path-pairs by type is shown in Fig. SM3a. Native PRNs have a different EDS path-pair distribution profile than non-native PRNs. In agreement with Figs. 2b & 2c, a significantly larger fraction of EDS path-pairs in native PRNs (6250) are edge-wise independent with zero stretch. Irrespective of edge independence, EDS path-pairs with positive stretch (type *01* or *11*) were more likely to traverse at least one short-cut that will disappear in the next step (Fig. 7a). The same affinity is observed with added short-cuts by path-pair type (Fig. 7b). Hence, path-pair stretch may be a more significant factor than edge independence in determining whether a short-cut edge will be deleted.

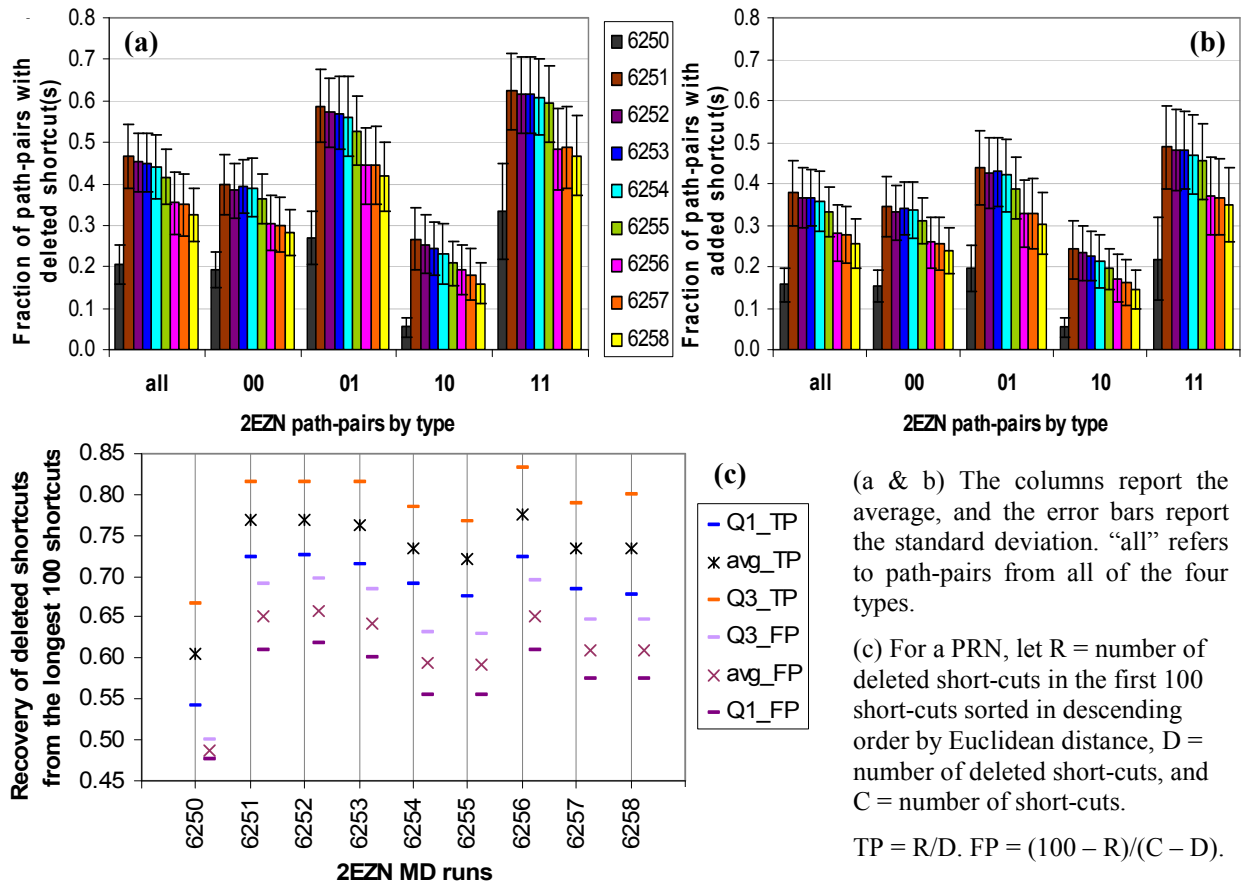

**Fig. 7 (a)** A significantly larger fraction of EDS path-pairs with positive stretch (*01* and *11*) traverse at least one deleted short-cut. **(b)** A significantly larger fraction of EDS path-pairs with positive stretch traverse at least one added short-cut. In (a & b), the fraction of path-pairs by type with at least one deleted (added) short-cut was calculated for each snapshot (PRN) in a run, and averaged over all snapshots for a run. **(c)** The average (avg) true positive rate (TP) and false positive rate (FP) of recovering deleted short-cuts from a PRN's short-cut set. Q1 and Q3 are the first and third quartiles respectively.

An average of 73.41% (std. dev. = 5.26%) of deleted short-cuts are recoverable in the first 100 short-cut edges sorted by Euclidean distance in descending order with an average false positive rate (FPR) of 61.08 % (std. dev. = 5.27%) (Fig. 7c). The recovery rate (TPR) is much better for non-native than native runs. TPR is much higher than expected by random chance which on average is 13.03% for 6250, and for the non-native runs 6251…6258, ranges from 19.15% to 32.17% (Fig. 6a). The FPR is high by conventional standards, but is a consequence of the number of edges required to create a spanning tree with 101 nodes (although non-native gSCNs may not span the entire PRN, Fig. 1b), and $|SC| \approx 2N$. It remains to be seen whether an even larger FPR is necessary given the transitivity of SCNs.

*SCN formation as a dynamic graph*
For 2EZN, almost all of the short-cuts deleted at step $t+1$ come from $gSCN_t$ and about 85% of the short-cuts added at step $t+1$ are edges in the set of all PRN edges with both endpoints in $gSCN_t$. This is expected from the definition of gSCN as the largest SCN component, and the property that gSCN spans almost all PRN nodes (Fig. 1b). It turns out that most deleted short-cuts (edges that under a *d21* or *d20* transition) are replaced by an edge (which becomes an added short-cut via a *d12* transition) in their cut-set with respect to a spanning tree. This replenishing process suggests a non-random relationship between deleted and added short-cuts, and together with strong SCN transitivity (Fig. 6), contributes to the growth of a gSCN. The idea of replacing a deleted edge with another edge from its cut-set to maintain connectivity in a graph comes from dynamic graph theory, and is applied to the SCN formation problem as follows:

1. Identify $delSC_{t+1}$, the set of deleted short-cuts at step $t+1$. An edge in $delSC_{t+1}$ is a short-cut edge at step $t$ and either a non-short-cut edge or a non-edge at step $t+1$. Deleted short-cuts are those that make either a *d21* or *d20* transition (Fig. 5).

2. Create a spanning tree $ST_t$ from $gSCN_t$, taking care to pack into $ST_t$ as many of the edges in $delSC_{t+1}$ as possible. We did this by first creating a forest of spanning trees[27] from $delSC_{t+1} \cap gSCN_t$ (the deleted short-cuts in the largest SCN component at step $t$), and then using the other short-cut edges in $gSCN_t$ to grow and join the trees to complete the construction of $ST_t$ (details in the Supplementary Material).

3. Generate the cut-set for each edge $e$ in $ST_t \cap delSC_{t+1}$. The removal of an edge $e$ from $ST$ splits $ST$ into two sub-trees. All PRN edges with one endpoint in one sub-tree and another endpoint in the other sub-tree complete the cut-set for $e$. $CUTS_t$ is the union of cut-sets for edges in $ST_t \cap delSC_{t+1}$.

4. Identify $addSC_{t+1}$, the set of added short-cuts at step $t+1$. An edge in $addSC_{t+1}$ is a non-short-cut edge at step $t$ and a short-cut edge at step $t+1$. Added short-cuts are edges that make a *d12* transition in Fig. 5. It is also possible for a non-edge at step $t$ to become a short-cut at step $t+1$ (via a *d02* transition) but such an added short-cut edge will not be in any cut-set generated in step 3. Non-edges that become short-cuts in the next step (*d02* edges) are much fewer than *d12* edges (Fig. SM4). Nonetheless, *d02* edges may be pivotal for correct SCN formation and this may be a limitation of our current approach.

5. Define $gSCN'_t$ as $gSCN_t$ augmented by all edges in $PRN_t$ that have both endpoints in $gSCN_t$. For each edge in $gSCN'_t \cap addSC_{t+1}$, find a cut-set from step 3 that includes it. If a cut-set is found, then we say that the added short-cut edge is *used* (as a replacement) and the deleted short-cut edge that generated the matching cut-set is *replaced*.

The fraction of deleted short-cuts that are not replaced (those without any added short-cuts in their cut-sets) and the fraction of unused added short-cuts (those that do not appear in any cut-set of a deleted short-cut) were calculated for each snapshot (PRN) and the average over all snapshots in a MD run are reported in Figs. 8 and 9 for the two MD datasets. The quality of $CUTS_t$ depends on the edges that make up $ST_t$. Better results (lower un-replaced and unused rates) are expected with larger $CUTS_t$. For this reason, we repeated the test on five spanning trees generated with a different random seed each time. The results from these different spanning trees (superimposed in Figs. 8 & 9) are not significantly different from each other.

For the 2EZN MD dataset, an average of 15.45% (std. dev. = 2.77%) of deleted short-cuts went un-replaced (*opt_st* in Fig. 8a), and an average of 11.41% (std. dev. = 4.13%) of added short-cuts were unused (*opt_st* in Fig. 8b). There may be some redundancies and a less than 100% success rate may provide sufficient constraints or even be necessary for SCN growth. This point needs to be tested and is discussed in the next section. The native state (6250) PRNs have significantly higher failure rates than the non-native state PRNs. This may be because our assumption that edge replacement is important for gSCN formation does not apply as well to already formed SCNs.

For the Villin MD dataset, an average of 19.41% (std. dev. = 1.76%) of deleted short-cuts went un-replaced (*opt_st* in Fig. 9a), and an average of 16.92% (std. dev. = 3.22%) of added short-cuts were unused (*opt_st* in Fig. 9b). Interestingly, the "successful" runs had a lower (p-value = 0.0398) average un-replaced rate, but a higher (p-value = 0.0339) average unused rate than the "unsuccessful" runs. These two factors could account for the larger SCNs in the "successful" runs (Table 1). The "unsuccessful" runs not only suffer from weaker SCN transitivity (Table 1), they also face a larger $q$ (probability of a short-cut edge being deleted in the next step) than the "successful" runs. The average $q$ for "unsuccessful" runs is 0.2226 (std. dev. 0.0204) and for "successful" runs is 0.1994 (std. dev. 0.0232). The smallest maximum $q$ over all runs is 0.4314, which suggests that gSCNs would need to be sufficiently robust to withstand the stress of edge mutations if they are to grow (Fig. 6).

When any edge in $gSCN'_t$ (including non-short-cut edges) can be used to grow and join the spanning trees in step 2, failure rates are significantly higher for both MD datasets. For the 2EZN MD dataset, an average of 16.55% (std. dev. = 3.89%) of deleted short-cuts were un-replaced (*ropt_st* in Fig. 8a), and an average of 18.84% (std. dev. = 4.77%) of added short-cuts were unused (*ropt_st* in Fig. 8b). For the Villin MD dataset, an average of 21.83% (std. dev. = 1.72%) of deleted short-cuts were un-replaced (*ropt_st* in Fig. 9a), and an average of 24.79% (std. dev. = 2.24%) of added short-cuts were unmatched (*ropt_st* in Fig. 9b). For both MD datasets, *ropt_st* produced significantly smaller $CUTS_t$ than *opt_st* (Figs. 8c & 9c), which accounts for the larger failure rates. The differences between "successful" and "unsuccessful" runs in terms of un-replaced and unused rates reported with *opt_st* were not significant with *ropt_st*. The results produced by *opt_st* in this analysis, which is significantly different from the outcomes produced by its randomized counterpart *ropt_st*, support the notions that (i) short-cuts are a set of distinctly placed edges within PRNs (as we have reported earlier [13]), and (ii) a non-random relationship exists between the deleted and added short-cut sets of a step.

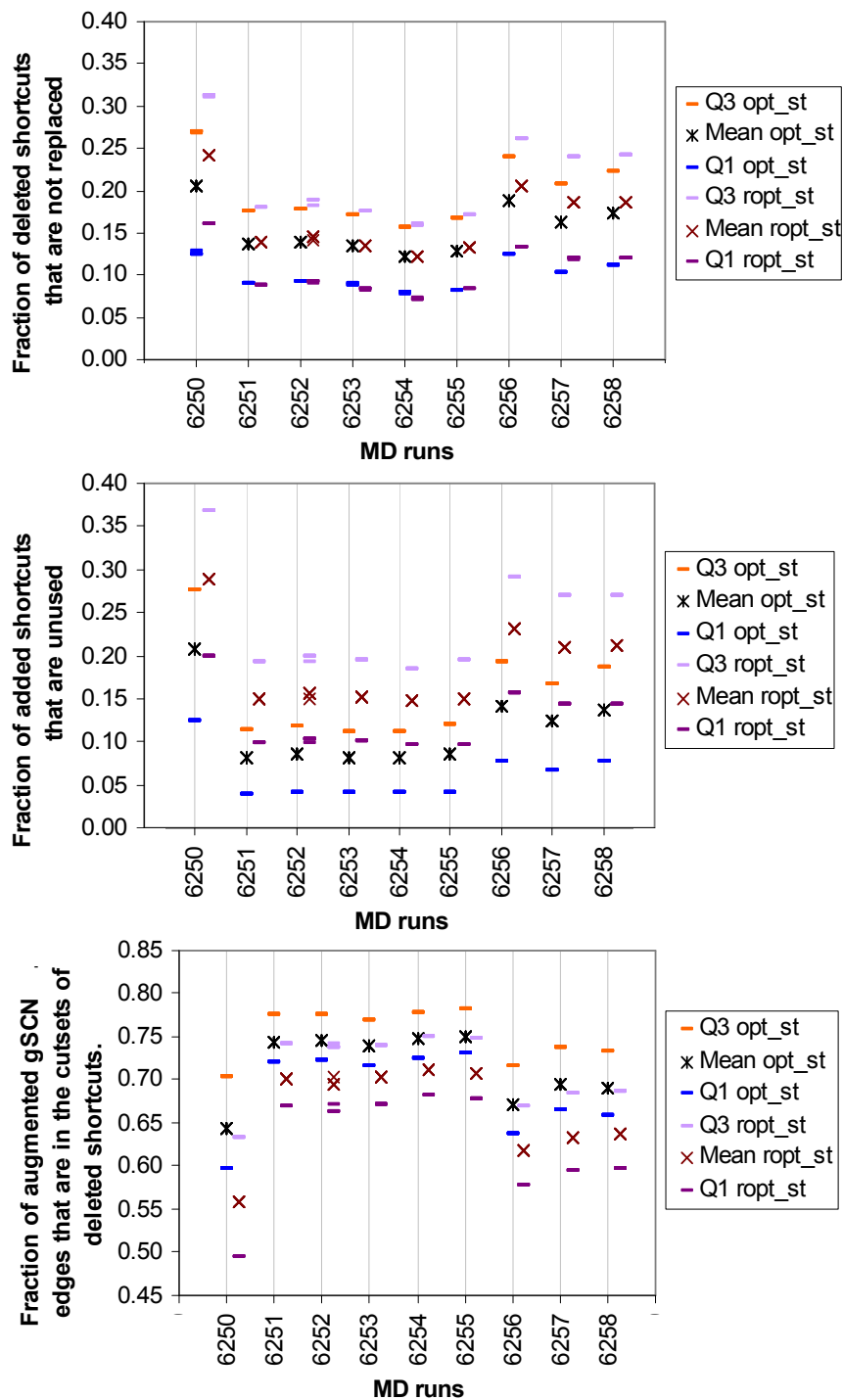

(a) A deleted short-cut is unreplaced if an added short-cut is not found in its cut-set w.r.t. a spanning tree *ST*.

Averaged over all runs, about 15.45% (std. dev. 2.77%) of deleted short-cuts went unreplaced (*opt_st*).

This percentage increased significantly to 16.55% (std. dev. 3.89%) when *ST* is supplemented with random edges in step 2 (*ropt_st*).

(b) An added short-cut is unused if it does not appear in the cut-set of any deleted short-cut w.r.t. a spanning tree *ST*.

Averaged over all runs, about 11.41% (std. dev. 4.13%) of added short-cuts go unused (*opt_st*).

This percentage increased significantly to 18.84% (std. dev. 4.77%) when *ST* is supplemented with random edges in step 2 (*ropt_st*).

(c) The fraction of edges in gSCN'$_t$ that are in CUTS$_t$ is significantly smaller when *ropt_st* is used to produce the edge cut-sets than when *opt_st* is used. CUTS$_t$ is the union of cut-sets for edges in $ST_t \cap$ delSC$_{t+1}$.

Accordingly, *ropt_st* has poorer outcomes in (a) & (b) above. That the *opt_st* and *ropt_st* outcomes are significantly different supports the notion that the short-cuts are a distinct set of PRN

**Fig. 8** Results for 2EZN showing the relationship between deleted and added short-cut sets.

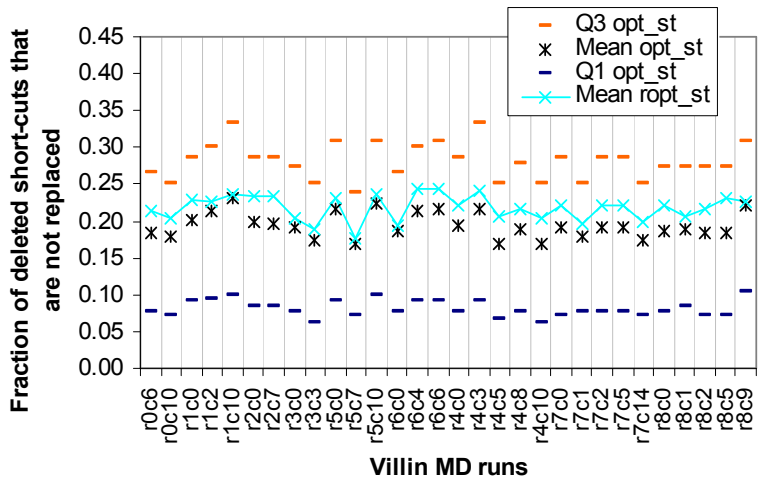

(a) Averaged over all clones, about 19.41% (std. dev. 1.76%) of deleted short-cuts were un-replaced (*opt_st*).

This percentage increased significantly to 21.83% (std. dev. 1.72%) when *ST* is supplemented with random edges in step 2 (*ropt_st*).

The successful runs have lower average un-replaced rate than the unsuccessful runs. This difference is not significant with *ropt_st*.

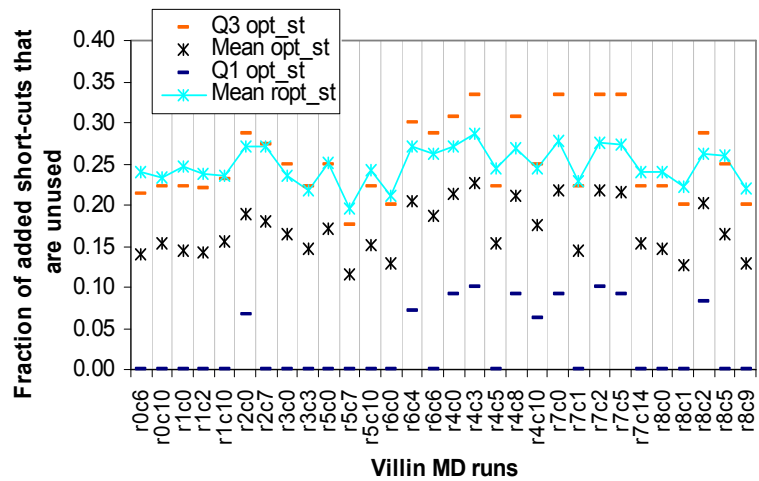

(b) Averaged over all clones, about 16.92% (std. dev. 3.22%) of added short-cuts are unused (*opt_st*).

This percentage increased significantly to 24.79% (std. dev. 2.24%) when *ST* is supplemented with random edges in step 2 (*ropt_st*).

Successful runs have a significantly higher average unused rate than unsuccessful runs. This difference is not significant with *ropt_st*.

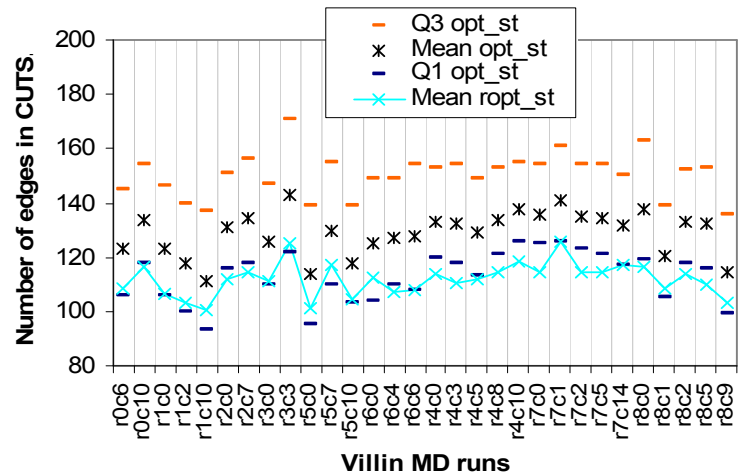

(c) $CUTS_t$ is the union of cut-sets for edges in $ST_t \cap delSC_{t+1}$.

The number of edges in $gSCN'_t$ that are covered by the cut-sets of deleted short-cuts is significantly smaller when random edges are used in step 2 (*ropt_st)* than when only short-cuts are used (*opt_st*).

Accordingly, *ropt_st* produced poorer outcomes in (a) & (b) above.

**Fig. 9** Results for HP-35 NLE NLE showing the relationship between deleted and added short-cut sets.

## Discussion

Our goal in describing the short-cut networks (SCNs) of protein residue networks (PRNs) is to obtain some guidelines for building a graph theoretic model of navigable SWN evolution within proteins. EDS paths may seem a little convoluted to construct, compared to BFS paths which is popular in PRN analysis. Nevertheless, due to its less diffusive nature, our investigations (section 3.7 in [34]) find that EDS paths have significantly better communication propensity [22, 28] than BFS paths. This property suggests that EDS paths are more representative of intra-protein communication than BFS paths.

While we have shown that SCNs can reflect differences in folding pathways within the framework model, our preliminary investigations did not distinguish the predicted folding pathways [25] as unique by any of our metrics introduced in this paper. The predicted folding pathways may be one of multiple feasible folding pathways, and a more detailed study of the search space is desirable. An enumeration of the search space may even be possible since SCNs and its properties are faster to compute then energy calculations.

Nonetheless, since the development of a well-formed SCN is observed to correlate strongly and positively with secondary structure formation, SCN formation is really more aligned with a protein folding model that does not assume the complete formation of secondary structure elements before higher level structures. Analogously, there are two ways to form (navigable) SWNs: (i) start with a regular graph and add randomness, or (ii) start with a random graph and add regularity. Both approaches have been probed extensively in the literature on graph theory and complex networks. The challenge is to apply these results to the protein folding problem.

SCNs, though a sub-network of PRNs, are not themselves navigable SWNs. Modeling the evolution of SCNs may require other computational tools to flesh out intermediate PRNs with probable non-SCN links. The approach by Green and Grant [29] may be helpful in this regard: for instance, in place of the LOS interactions, short-cuts could be used as restraints to the CNS program.

We have thus far examined *atemporal* network statistics (measurements that do not involve time as a component) of sequences of static PRNs (snapshots). A logical extension of this work is to utilize the formalisms and methods emerging from the currently active field of time varying graphs [30] and dynamic networks, and examine temporal aspects of PRNs as a protein folds. By connecting the static and the dynamic aspects of PRNs within a temporal framework, the protein folding process becomes a problem of graph or path-set evolution with the purpose of forming suitable pathways within native proteins. Understanding the formation of the navigable SWNs within PRNs via SCNs may also elucidate the evolution of other networks since navigable SWNs are ubiquitous in natural and artificial systems.

## Methods

*PRN construction*

The PRNs are constructed as before [13] from the PDB coordinates files. Each amino acid (residue) is represented by a node and two nodes *u* and *v* are linked *iff* $|u - v| \geq 2$ and their interaction strength $I_{uv}$ is

≥ 50%. $I_{uv} = \frac{n_{uv} \times 100}{\sqrt{N_u \times N_v}}$ where $n_{uv}$ is the number of distinct atom-pairs $(I, j)$ such that $I$ is an atom of residue $u$, $j$ is an atom of residue $v$ and the Euclidean distance between atoms $I$ and $j$ is ≤ 7.5 Å. $N_u$ and $N_v$ are normalization values by residue types [31].

*The EDS algorithm and short-cut edges*

EDS is a greedy local search algorithm, similar to Kleinberg's [14], but with backtracking capability. At each step of a search, EDS surveys the proximity to target of the current node's direct neighbors in the PRN, and moves to an as yet unvisited node $x$ which is closest (Euclidean distance) amongst all nodes surveyed so far, to a target node. EDS remembers proximity information from all nodes surveyed so far in a search, but EDS's memory is short-term – proximity information is gathered afresh with each search. So it is possible that $x$ is not adjacent to the current node. In this case, EDS retraces its steps (*backtrack*) until $x$ becomes reachable. For example, in the search for target node 75 from source node 62 (Fig. 10 right), EDS finds after visiting node 76 that node 77 is the closest as yet unvisited node to node 75, and has to retrace edge (74, 76) to reach node 77. An edge $(u, v)$ is a *short-cut* if and only if $L^T(v) = L^T(u) + 1$, and $v$ is adjacent to a node $w$ such that $L^T(w) < L^T(u)$. $L^T(x)$ is a positive integer denoting the *order* EDS visits nodes in a path $T$ for the first time.

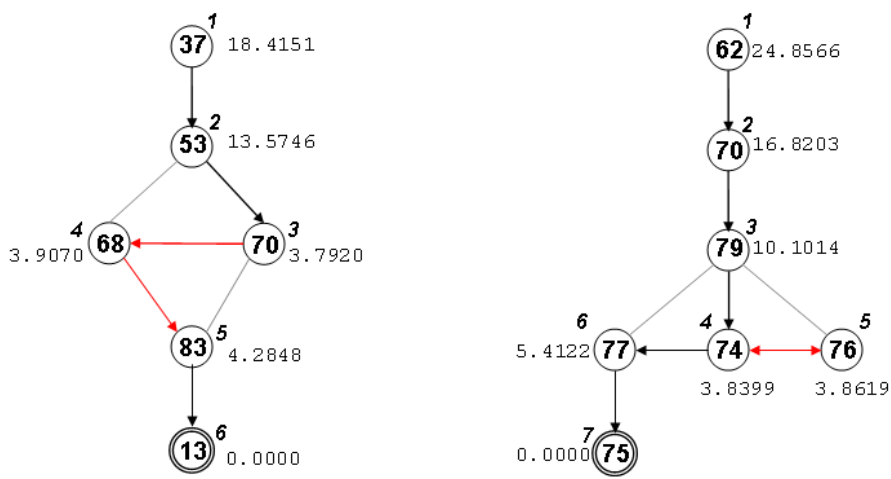

**Fig. 10** Two EDS paths from a 2EZN PRN (run 6253 frame 20). The EDS path on the left is of length five and is ⟨37, 53, 70, 68, 83, 13⟩. The EDS path on the right is of length seven and is ⟨62, 70, 79, 74, 76, 74, 77, 75⟩. PRN edges are undirected, but the edges are oriented in the diagram in the direction they are traversed by EDS in the respective paths. Un-oriented edges are not traversed, but exist and play a role in determining whether en edge is a short-cut. Short-cut edges are marked in red. Bi-directional edges are retraced/backtrack edges. The real number besides each node is the node's Euclidean distance to the target node. The italicized integer besides each node $x$ is the node's first visit order, $L^T(x)$.

*SCN and its random counterparts*

Define *SC* as the set of short-cut edges identified by EDS for a PRN. The short-cut network (*SCN*) of a PRN is a graph whose edge set is *SC*, and whose vertex set is induced by its edge set. For comparison, we constructed two types of pseudo short-cut networks: random short-cut (*rsc*) and random spanning tree (*rst*). A *rsc* network is constructed from |*SC*| PRN edges chosen uniformly at random. A *rst* network is also constructed from |*SC*| PRN edges but the selected edges come from the union of two or more trees spanning the PRN found via random walks. *rst* networks are splicers [32] and are included in this work because |*SC*| ≈ 2*N*. Multiple *rsc* and *rst* networks are generated per SCN with a different random seed each time. Usually, we are interested in a SCN's largest connected component, denoted gSCN. Analogously, *grsc* (*grst*) denotes the largest connected component of a *rsc* (*rst*) network. The existence of a *grsc* (*grst*) is expected since with |*SC*| ≈ 2*N*, the average degree of a *rsc* (*rst*) network is approximately 2(2*N*)/*N* = 4, which is larger than the threshold required for a random graph to have a giant component [33].

*Molecular Dynamics (MD) datasets*

The 2EZN MD dataset [16-18] has nine runs, each with a different number of snapshots. The runs simulate the native (6250) and unfolding (6251…6258) dynamics of the mainly beta-sheet 2EZN protein which has 101 amino acids. The Villin MD dataset, which simulates the folding of a subdomain of the mainly alpha-helix 2F4K protein from different denatured configurations, is publically available from https://simtk.org/home/foldvillin. The subdomain consists of 35 amino acids, and is modified for fast folding (HP-35 Nle Nle). The MD runs in this dataset is partitioned into two sets [19]. In the "successful" set are runs 4, 7, and 8. The starting structures of these runs either folded much faster or folded to a significant extent. In the "unsuccessful" set are runs 0, 1, 2, 3, 5, and 6. The starting structures of these runs generated trajectories that either only briefly visited a folded configuration or did not fold at all. We chose at least two trajectories at random from each run, with preference for those with more frames (a frame comprises many shapshots). For both datasets, a PRN is constructed for every snapshot or frame, and the majority of the MD results we report are produced by averaging a statistic over all snapshots in a run.

## Acknowledgements

This work was made possible by the facilities of the Shared Hierarchical Academic Research Computing Network (SHARCNET:www.sharcnet.ca) and Compute/Calcul Canada. Funding was provided in part through a post-doctoral research position at Memorial University. Thanks the Dynameomics group for allowing access to the 2EZN MD data.

## Supplementary information

*Notes on building the ST spanning trees to match deleted with added short-cuts.*

We adopted Kruskal's approach [27] of constructing spanning trees in step 2. We found that starting with a forest of trees made up of deleted short-cuts worked better (yielded lower un-replaced and unused rates) for both MD datasets than growing the spanning tree as a single tree starting from a single edge. Some deleted short-cuts in a gSCN will still be left out of *ST*, but this is unavoidable since SCNs are strongly transitive. Cycles of deleted short-cuts are broken up by removing an edge chosen at random from the cycle, and the analysis is repeated on five *ST*s constructed with a different random number seed each time. Results from the different *ST*s did not vary significantly from each other.

For an added short-cut *f* to be a replacement edge of a deleted short-cut *e* = (*u*, *v*) with respect to a spanning tree *ST*, *e* must be part of *ST*, and *f* needs to "straddle" *e*, i.e. *f* must be part of some path from *u* to *v* in gSCN′ that does not involve *e*. To increase the chance of this happening, we try to place nodes incident to many added short-cuts on the periphery of *ST*. Observe that leaf nodes in a spanning tree have more possible edge partners than non-leaf nodes, and edges between nodes in the same sub-tree cannot contribute to a cut-set.

**2F4K (Native, PDB structure)**

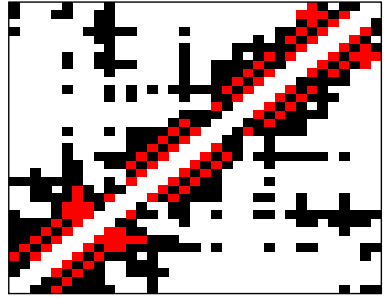

**Run0 ("unsuccessful")**       **Run1 ("unsuccessful")**       **Run2 ("unsuccessful")**

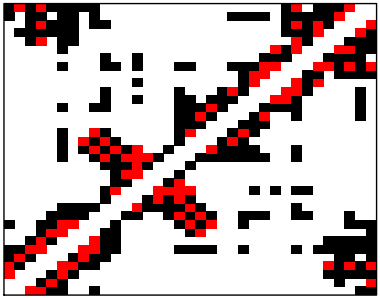 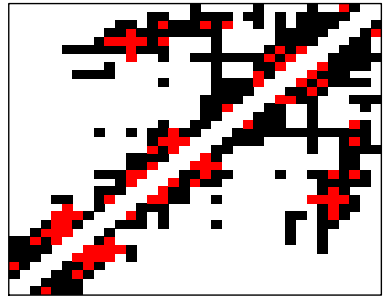 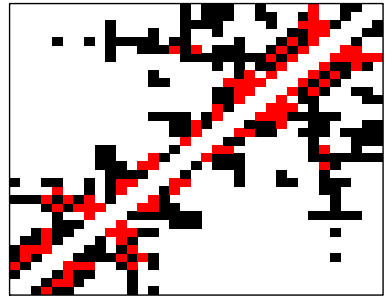

**Run3 ("unsuccessful")**       **Run5 ("unsuccessful")**       **Run6 ("unsuccessful")**

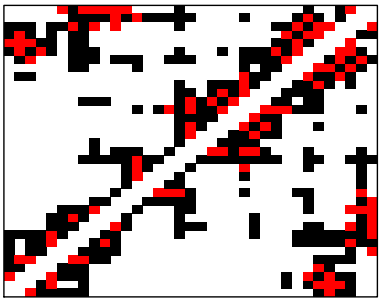 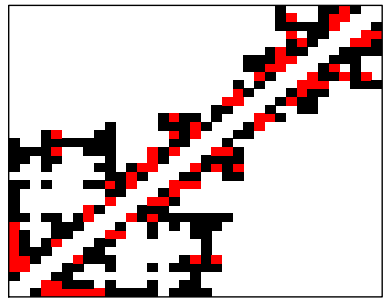 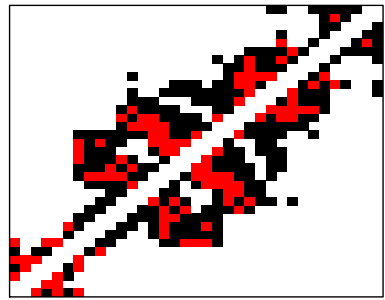

**Run4 ("successful")**       **Run7 ("successful")**       **Run8-9 ("successful")**

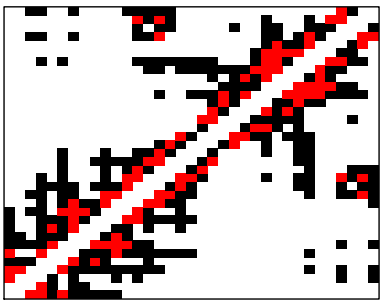 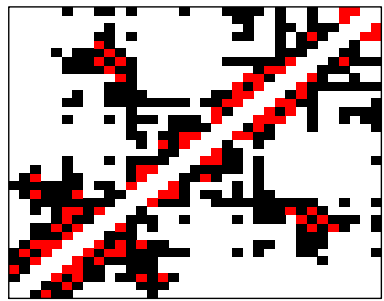 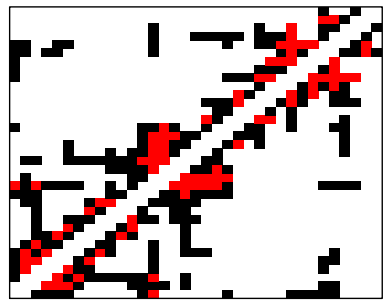

**Fig. SM1** Native and initial (starting structure) Villin PRNs and SCNs. Each plot is a 35 x 35 contact map (adjacency matrix). A red cell ($x$, $y$) denotes a short-cut edge between nodes $x$ and $y$. A black cell denotes a non-short-cut edge.

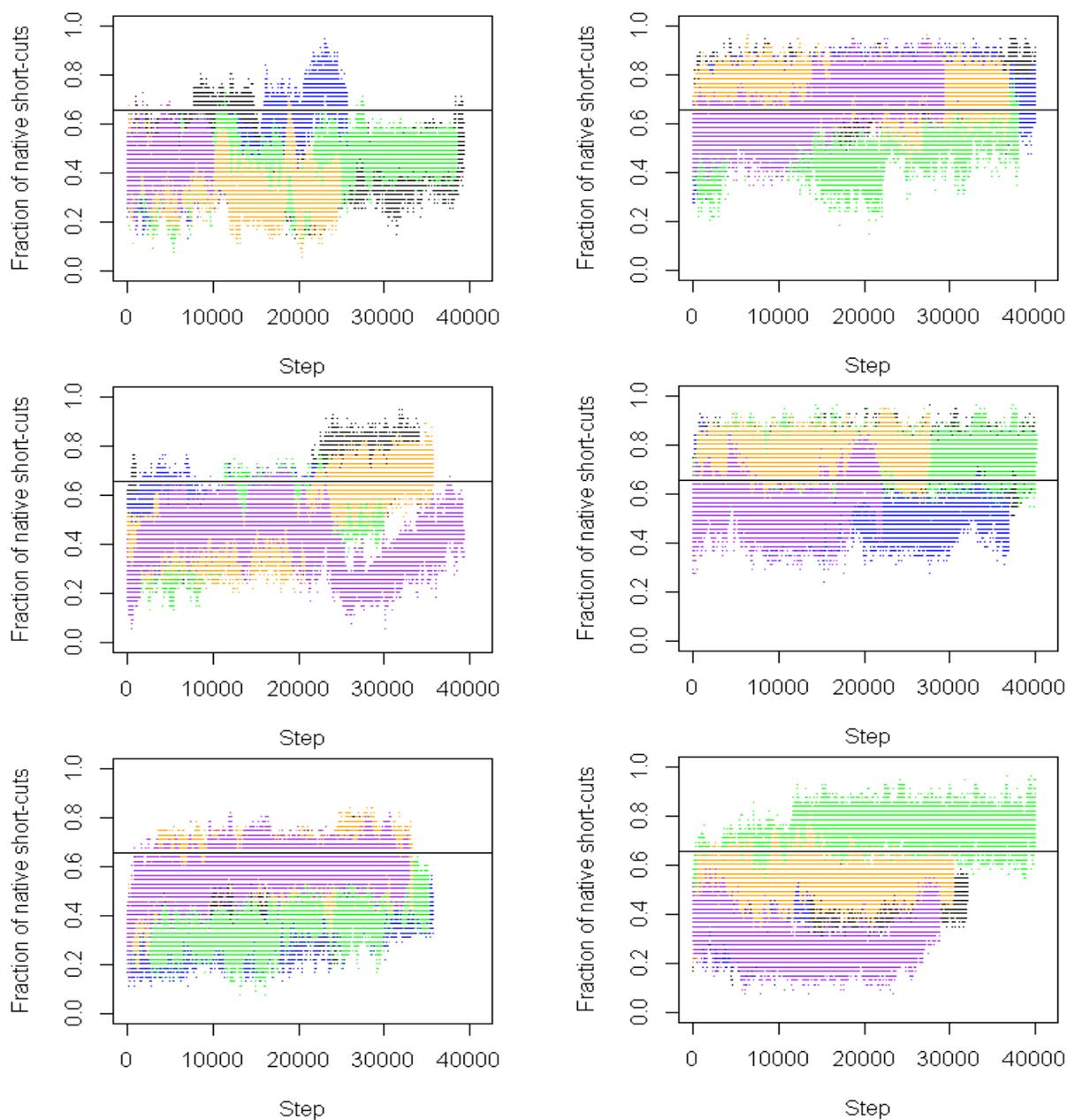

**Fig. SM2** Fraction of native short-cuts is the number of native short-cuts per SCN at step *t* divided by the number of native short-cuts, which is 55. Each plot shows the fraction of native short-cuts at each step for five Villin MD runs. Plots on the left are of the "unsuccessful" trajectories, which from top to bottom are: r0c6 (6.02%), r0c10 (20.82%), r1c0 (0.18%), r1c2 (0.12%), r1c10 (0.38%), r2c0 (33.99%), r2c7 (4.04%), r3c0 (4.37%), r3c3 (24.11%), r5c0 (0.69%), r5c7 (3.02%), r5c10 (0.00%), r6c0 (0.08%), r6c4 (42.16%) and r6c6 (22.69%). The fraction of snapshots in a trajectory with more than 35 native short-cuts is given as a percentage. Plots on the right are of the "successful" trajectories, which from top to bottom are: r4c0 (90.33%), r4c3 (90.79%), r4c5 (11.62%), r4c8 (91.29%), r4c10 (53.82%), r7c0 (95.59%), r7c1 (15.61%), r7c2 (93.95%), r7c5 (94.23%), r7c14 (15.99%), r8c0 (0.28%), r8c1 (0.02%), r8c2 (76.90%), r8c5 (3.39%) and r8c9 (0%). The horizontal line marks y=36/55=0.6545. Compared to the "unsuccessful" runs, the "successful" runs, in particular those from r4 (top right) and r7 (middle right), spend more time above the horizontal line. The only trajectory from r8 (bottom right) that does this is r8c2 (green). Formation of the native SCN seems to be a common attribute of "successful" runs that is uncommon in the "unsuccessful" runs. Trajectories from r8 are of interest since they all have the same starting configuration (Run8-9), but produce distinctly different trajectories.

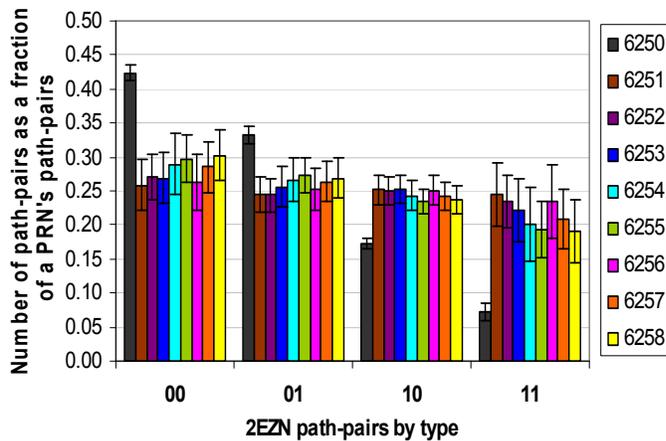

**Fig. SM3** Distribution of path-pairs, with more than one edge, by type. The distribution profile of the native run (6250) is distinct from the other non-native runs.

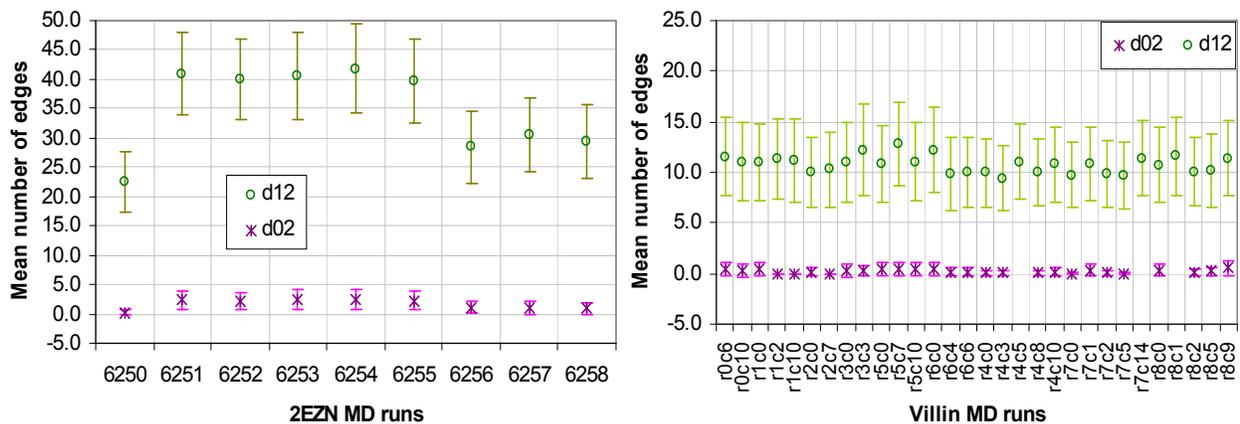

**Fig. SM4** The average number of short-cuts added in a step for the two MD datasets. *d12* denotes non-short-cut edges that become short-cuts in the next step/snapshot. *d02* denotes non-edges that become short-cut edges in the next step. *d02* edges are much fewer than *d12*, and are excluded by our matching algorithm. Error bars denote standard deviation about the mean.